\documentclass{article}

\usepackage[preprint]{neurips_2025}

\usepackage[T1]{fontenc}
\usepackage[utf8]{inputenc}
\usepackage{microtype}
\usepackage{graphicx}
\usepackage{booktabs}
\usepackage{amsmath,amssymb,mathtools}
\usepackage{enumitem}
\usepackage{algorithm}
\usepackage{algorithmic}
\usepackage{url}

\title{Modular Layout Synthesis (MLS): Front-end Code via Structure Normalization and Constrained Generation}

\author{
  Chong Liu, Ming Zhang, Fei Li, Hao Zhou, Xiaoshuang Chen, Ye Yuan\\
  Nanjing University \\
}

\begin{document}
\maketitle

\begin{abstract}
Automated front-end engineering drastically reduces development cycles and minimizes manual coding overhead.
While Generative AI has shown promise in translating designs to code, current solutions often produce monolithic scripts, failing to natively support modern ecosystems like React, Vue, or Angular.
Furthermore, the generated code frequently suffers from poor modularity, making it difficult to maintain.
To bridge this gap, we introduce \textbf{Modular Layout Synthesis (MLS)}, a hierarchical framework that merges visual understanding with structural normalization.
Initially, a visual-semantic encoder maps the screen capture into a serialized tree topology, capturing the essential layout hierarchy.
Instead of simple parsing, we apply heuristic deduplication and pattern recognition to isolate reusable blocks, creating a framework-agnostic schema.
Finally, a constraint-based generation protocol guides the LLM to synthesize production-ready code with strict typing and component props.
Evaluations show that MLS significantly outperforms existing baselines, ensuring superior code reusability and structural integrity across multiple frameworks.
\end{abstract}

\section{Introduction}

\paragraph{Background.}
Front-end user interface (UI) development remains one of the most time-consuming stages of modern software delivery: developers translate pixel-level designs into hierarchical layouts, pick framework components, and iteratively debug style and structure.
Recent multimodal large language models (MLLMs) and vision-language models (VLMs) have made UI-to-Code feasible by conditioning code generation on screenshots or design mockups \cite{beltramelli2018pix2code,si2025design2code,laurencon2024websight}.
However, a key practical requirement in real engineering settings is \emph{framework-awareness}: production UIs are rarely written as plain HTML/CSS only, but instead rely on component-based ecosystems such as React, Vue, and Angular.

\paragraph{Problem.}
Despite strong progress, existing UI-to-Code systems often produce \emph{monolithic} outputs: long flat files where repeated patterns are expanded verbatim, component boundaries are implicit, and reusability is poor \cite{si2025design2code,wan2025dcgen}.
Such code is hard to maintain (high duplication, weak typing), hard to reuse (no clean props or loops), and hard to port across frameworks (syntax differences leak throughout the output).
Empirically, even when the rendered UI is visually close, the resulting code can have low structural quality and low component-level reuse.

\paragraph{Prior work.}
Prior directions include (i) end-to-end screenshot-to-code models \cite{beltramelli2018pix2code,lee2023pix2struct,laurencon2024websight},
(ii) prompt-based or divide-and-conquer pipelines \cite{wan2025dcgen},
(iii) layout-guided code generation \cite{wu2025layoutcoder},
and (iv) benchmarks and evaluation protocols for UI2Code \cite{si2025design2code,gui2024vision2ui,xu2025webvia}.
However, these methods typically couple \emph{layout understanding}, \emph{reuse discovery}, and \emph{framework-specific code emission} in a single step, making it difficult to guarantee modular outputs or portability.

\paragraph{Our solution.}
We propose \textbf{Modular Layout Synthesis (MLS)}, a hierarchical framework that explicitly decomposes UI-to-Code into three stages:
(1) \emph{visual-to-structure}---a lightweight visual-semantic encoder predicts a coarse DOM-like tree in a constrained serialized format;
(2) \emph{structure normalization and component mining}---deterministic canonicalization and pattern mining isolate reusable blocks and repeated structures into a framework-agnostic intermediate representation (IR);
(3) \emph{constraint-based multi-framework generation}---a framework-conditioned protocol guides an LLM to emit typed components, props, and repeat constructs without expanding duplicates.
We summarize our contributions:
\begin{itemize}[leftmargin=1.2em]
\item We introduce MLS, a modular UI-to-Code framework that separates \emph{layout parsing}, \emph{reuse extraction}, and \emph{framework-specific synthesis}.
\item We propose a normalization and component mining procedure (hashing + clustering + conflict resolution) that yields a portable IR capturing components, repeats, and typed props.
\item We design a constraint-based generation protocol enabling consistent code structure across HTML/React/Vue/Angular, improving portability and reusability in experiments.
\end{itemize}

\section{Related Work}
We discuss four threads of research.

\paragraph{UI-to-Code and design-to-code.}
Early work explored end-to-end generation from GUI screenshots \cite{beltramelli2018pix2code}, while later systems scaled data and evaluation \cite{si2025design2code,laurencon2024websight}.
Recent pipelines incorporate prompting, segmentation, or iterative refinement \cite{wan2025dcgen,xu2025webvia}.
Datasets such as RICO \cite{deka2017rico} and Vision2UI \cite{gui2024vision2ui} support layout-aware modeling.

\paragraph{Screenshot parsing and markup understanding.}
Parsing visually rich screenshots into structured representations is a common pretraining signal \cite{lee2023pix2struct}.
Document/layout models such as LayoutLM series \cite{xu2020layoutlm,xu2021layoutlmv2,huang2022layoutlmv3} and OCR-free models \cite{kim2022donut} motivate robust visual-to-structure mappings.
Markup-aware pretraining (e.g., HTML/XML) also improves structural reasoning \cite{li2022markuplm}.

\paragraph{Code LMs and constrained decoding.}
Code generation benefits from code-specialized LMs \cite{wang2021codet5,nijkamp2022codegen,fried2022incoder}.
To improve syntactic correctness, grammar-constrained decoding enforces CFG constraints during generation \cite{geng2023gcd,geng2023flexible}, while grammar-aligned decoding studies quality distortions \cite{park2024gad}.
Beyond syntax, type-constrained decoding can further reduce compilation errors \cite{mundler2025typeconstrained}.
Structured generation frameworks and evaluations are increasingly standardized around JSON-schema-like constraints \cite{structuredoutputs2024openai,structuredgen2025}.

\section{Approach: Modular Layout Synthesis via Blueprint Compression}

\subsection{Notation and Output Contract}
Let the input screenshot be $s \in \mathbb{R}^{H \times W \times 3}$ and the target runtime be
$\mu \in \{\textsc{HTML},\textsc{React},\textsc{Vue},\textsc{Angular}\}$.
Our goal is to generate a code bundle $\mathcal{Z}_\mu$ that (i) visually matches $s$ after rendering,
(ii) exposes explicit reusable components with typed properties, and
(iii) preserves repeated structures as loop constructs rather than expanded copies.

MLS is implemented as a three-module factorization:
\begin{equation}
\mathcal{A} = \Phi_{\alpha}(s), \qquad
\Omega = \mathfrak{C}(\mathcal{A}), \qquad
\mathcal{Z}_\mu = \Psi_{\beta}(\Omega, \mu),
\label{eq:mls_factorization_new}
\end{equation}
where $\Phi_\alpha$ produces a \emph{layout arborescence} $\mathcal{A}$, $\mathfrak{C}$ compresses it into a \emph{portable blueprint} $\Omega$ with motifs and typed fields, and $\Psi_\beta$ realizes framework-specific code under hard constraints.

\paragraph{Layout arborescence.}
We represent structure as a rooted ordered arborescence
$\mathcal{A}=(\mathcal{V},\mathcal{E},\varrho)$.
Each node $u\in\mathcal{V}$ carries a coarse category $c(u)\in\mathcal{G}$
and an optional geometry anchor $g(u)\in[0,1]^4$ (normalized $(x_0,y_0,x_1,y_1)$):
\begin{equation}
u \equiv \big(c(u),\, g(u),\, \kappa(u)\big),
\end{equation}
where $\kappa(u)$ is a payload placeholder (text/image/link/input etc.) stored out-of-band (Module B).

\paragraph{Portable blueprint.}
The intermediate object is
\begin{equation}
\Omega \triangleq \big(\mathcal{S},\, \mathbb{M},\, \mathbf{X},\, \mathbb{T}\big),
\end{equation}
where $\mathcal{S}$ is a rewritten skeleton tree, $\mathbb{M}$ is a library of mined motifs (reusable templates),
$\mathbf{X}$ is a table (or a set of tables) of instance fields for loops/components, and $\mathbb{T}$ is a typing environment
(e.g., prop schemas and attribute types).

\subsection{Module A: Screen-to-Arborescence Transduction}
Module A predicts $\mathcal{A}$ from screenshot $s$ using a vision-semantic encoder-decoder.

\paragraph{Constrained serialization.}
We encode $\mathcal{A}$ into a bracketed topology string $\operatorname{br}(\mathcal{A})$
that is guaranteed to parse under a simple grammar (balanced braces, fixed keys).
The generator defines an autoregressive distribution:
\begin{equation}
p_\alpha\big(\operatorname{br}(\mathcal{A}) \mid s\big)
= \prod_{t=1}^{T} p_\alpha\big(w_t \mid w_{<t}, s\big),
\end{equation}
where $w_t$ is the $t$-th token of the serialization.

\paragraph{Geometry-aware objective.}
Given training pair $(s,\mathcal{A}^\star)$, we minimize
\begin{equation}
\mathcal{J}_{\mathrm{A}}(\alpha)
= \underbrace{\sum_{t=1}^{T^\star} \big(-\log p_\alpha(w_t^\star \mid w_{<t}^\star, s)\big)}_{\text{topology token loss}}
+ \lambda_g \underbrace{\sum_{u\in\mathcal{V}^\star}\ell_{\mathrm{box}}(g(u),g^\star(u))}_{\text{geometry loss}}
+ \lambda_r \underbrace{\sum_{u\in\mathcal{V}} \mathbf{1}\{\mathrm{area}(g(u))<\epsilon\}}_{\text{noise regularizer}},
\label{eq:moduleA_loss_new}
\end{equation}
where $\ell_{\mathrm{box}}$ can be $\ell_1$ or $(1-\mathrm{IoU})$, and the last term discourages tiny spurious nodes.

\paragraph{Stability constraints.}
To make the downstream deterministic compression robust, we enforce a coarse vocabulary:
\begin{equation}
\mathcal{G}=\{\texttt{wrap},\texttt{row},\texttt{col},\texttt{text},\texttt{media},\texttt{ctl},\texttt{link}\}.
\end{equation}
We also prune by depth and area:
\begin{equation}
\mathcal{V} \leftarrow \{u\in\mathcal{V}:\ \mathrm{depth}(u)\le d_{\max}\ \wedge\ \mathrm{area}(g(u))\ge a_{\min}\}.
\label{eq:prune_new}
\end{equation}

\subsection{Module B: Blueprint Compression with Motif Harvesting}
Module B converts $\mathcal{A}$ into a framework-agnostic blueprint $\Omega$ by (i) canonicalizing order and payload,
(ii) discovering repeated substructures (motifs), and (iii) inducing typed component schemas.

\subsubsection{B.1 Canonical skeletonization and payload ledger}
We split structure vs. content via
\begin{equation}
(\hat{\mathcal{A}},\ \mathcal{L}) = \operatorname{Skel}(\mathcal{A}),
\end{equation}
where $\hat{\mathcal{A}}$ is a \emph{skeleton} with all literal payloads replaced by typed holes,
and $\mathcal{L}$ is a \emph{payload ledger} that stores the removed literals keyed by canonical paths.

\paragraph{Canonical sibling order.}
For a node $u$ with children $(v_1,\dots,v_m)$, we reorder children by reading order:
\begin{equation}
(v_1,\dots,v_m) \leftarrow \operatorname{sort}\Big((v_1,\dots,v_m);\ \big(y_c(g(v)),x_c(g(v))\big)\Big),
\end{equation}
where $(x_c(\cdot),y_c(\cdot))$ is the box center. This ensures deterministic serialization for isomorphic subtrees.

\paragraph{Ledger addressing.}
Let $\chi(u)$ be the canonical path of node $u$ after ordering (sequence of child indices).
We store payload attributes (e.g., text string, src, href, placeholder):
\begin{equation}
\mathcal{L}[\chi(u)] \triangleq \mathrm{payload}(u).
\end{equation}

\subsubsection{B.2 Motif fingerprints and near-match unification}
We compute bottom-up fingerprints on the skeleton $\hat{\mathcal{A}}$.
Let $\zeta(u)$ be a normalized node label (category + role tag):
\begin{equation}
\zeta(u) \triangleq \big(c(u),\ \rho(u)\big).
\end{equation}
Define the digest $d(u)$ recursively:
\begin{equation}
d(u) = \mathcal{H}\!\left(\zeta(u)\ \Vert\ d(\mathrm{ch}_1(u))\ \Vert\ \cdots\ \Vert\ d(\mathrm{ch}_m(u))\right),
\label{eq:digest_new}
\end{equation}
where $\mathcal{H}$ is a fixed hash, and $\mathrm{ch}_i(u)$ are children after canonical sort.

\paragraph{Near-match signature.}
Exact hashes miss “almost same” motifs (e.g., extra icon node). We therefore compute a lightweight signature
$\sigma(u)\in\mathbb{R}^p$ as a normalized histogram of labels in the subtree:
\begin{equation}
\sigma(u) = \frac{1}{Z(u)}\sum_{w\in \mathrm{sub}(u)} \mathbf{e}_{\zeta(w)},
\qquad Z(u)=|\mathrm{sub}(u)|.
\end{equation}
We merge motifs if cosine similarity exceeds threshold $\eta$:
\begin{equation}
\mathrm{sim}(u,v) = \frac{\langle \sigma(u),\sigma(v)\rangle}{\|\sigma(u)\|_2\|\sigma(v)\|_2} \ge \eta.
\label{eq:nearsim_new}
\end{equation}

\subsubsection{B.3 Template induction and typed field schemas}
A motif cluster $\mathcal{Q}=\{u_1,\dots,u_k\}$ defines a reusable template $\mathfrak{m}_{\mathcal{Q}}$.
We create \emph{holes} for varying payload positions and produce an instance table $\mathbf{X}_{\mathcal{Q}}$:

\paragraph{Hole collection.}
Let $\mathcal{Holes}(\mathcal{Q})$ be the set of ledger keys inside the motif where values differ across instances.
For each hole $h\in\mathcal{Holes}(\mathcal{Q})$, we gather observed values
\begin{equation}
\mathcal{U}_h = \{\mathcal{L}[\chi_h(u_i)]\}_{i=1}^{k}.
\end{equation}

\paragraph{Deterministic type assignment.}
We infer a field type via a rule-based classifier $\operatorname{Type}(\cdot)$:
\begin{equation}
\tau(h) = \operatorname{Type}(\mathcal{U}_h)\in
\{\texttt{Str},\texttt{Num},\texttt{Bool},\texttt{Url},\texttt{Enum}\}.
\label{eq:type_new}
\end{equation}
This induces a prop schema for the motif:
\begin{equation}
\mathbb{T}(\mathfrak{m}_{\mathcal{Q}}) = \{(h,\tau(h)):\ h\in \mathcal{Holes}(\mathcal{Q})\}.
\end{equation}

\paragraph{Instance table.}
We pack per-instance values into a table $\mathbf{X}_{\mathcal{Q}}\in \mathbb{D}^{k\times |\mathcal{Holes}|}$:
\begin{equation}
\mathbf{X}_{\mathcal{Q}}[i,h] = \mathcal{L}[\chi_h(u_i)].
\end{equation}
The complete motif library is $\mathbb{M}=\{\mathfrak{m}_j\}$ and tables $\mathbf{X}=\{\mathbf{X}_j\}$.

\subsubsection{B.4 Non-overlap selection as weighted packing}
Motifs may overlap (nested candidates). We select a consistent set by weighted packing.
Each candidate motif $j$ has occurrences $\mathcal{O}_j$ and a size $\mathrm{sz}(j)=|V(\mathfrak{m}_j)|$.
We define a compression score:
\begin{equation}
w(j) \triangleq (|\mathcal{O}_j|-1)\cdot \mathrm{sz}(j) - \gamma,
\label{eq:weight_new}
\end{equation}
and solve the set packing (NP-hard) approximately by greedy:
\begin{equation}
\max_{\mathbf{z}\in\{0,1\}^J} \sum_{j=1}^{J} w(j)z_j
\quad \text{s.t.}\quad
z_j+z_{j'}\le 1\ \text{if}\ \mathcal{O}_j \cap \mathcal{O}_{j'}\neq\emptyset.
\label{eq:packing_new}
\end{equation}
We sort candidates by $w(j)$ and insert if it does not conflict, yielding the final skeleton $\mathcal{S}$ with
\texttt{macro} and \texttt{loop} nodes.

\begin{algorithm}[t]
\caption{Blueprint Compression $\mathfrak{C}(\mathcal{A})$ (Module B)}
\label{alg:compress_new}
\begin{algorithmic}[1]
\STATE $(\hat{\mathcal{A}},\mathcal{L}) \leftarrow \operatorname{Skel}(\mathcal{A})$
\STATE canonical-sort siblings by box-center order
\STATE compute digests $d(\cdot)$ by Eq.~\eqref{eq:digest_new}
\STATE build clusters by exact $d(\cdot)$, then merge by Eq.~\eqref{eq:nearsim_new}
\STATE induce motifs $\mathbb{M}$ and tables $\mathbf{X}$; infer types via Eq.~\eqref{eq:type_new}
\STATE score candidates by Eq.~\eqref{eq:weight_new} and greedily solve Eq.~\eqref{eq:packing_new}
\STATE rewrite skeleton into $\mathcal{S}$ with \texttt{macro}/\texttt{loop}; build typing env $\mathbb{T}$
\RETURN $\Omega=(\mathcal{S},\mathbb{M},\mathbf{X},\mathbb{T})$
\end{algorithmic}
\end{algorithm}

\subsection{Module C: Constraint-Driven Multi-Framework Realization}
Module C realizes $\Omega$ into framework-specific code $\mathcal{Z}_\mu$ using an LLM under a hard protocol.

\subsubsection{C.1 Realization map}
We define a deterministic realization map $\pi_\mu$ that translates blueprint nodes to framework constructs:
\begin{equation}
\pi_\mu:\ \{\texttt{wrap,row,col,text,media,ctl,link,macro,loop}\} \rightarrow \Lambda_\mu.
\end{equation}
In particular, loops are realized natively:
\[
\texttt{loop} \Rightarrow
\begin{cases}
\texttt{\{items.map((it)=\!>\!\dots)\}}, & \mu=\textsc{React},\\
\texttt{v-for="it in items"}, & \mu=\textsc{Vue},\\
\texttt{*ngFor="let it of items"}, & \mu=\textsc{Angular},\\
\texttt{(fallback) replicate nodes}, & \mu=\textsc{HTML}.
\end{cases}
\]

\subsubsection{C.2 Hard constraints as admissible-token filtering}
Let the LLM emit a token sequence $y_{1:T}$ describing a set of files $\mathcal{Z}_\mu$ under a file-block protocol.
At decoding step $t$, we restrict tokens to an admissible set:
\begin{equation}
\mathcal{V}_t
= \mathcal{V}^{\text{gram}}_t \cap \mathcal{V}^{\text{bind}}_t \cap \mathcal{V}^{\text{type}}_t,
\label{eq:admissible_new}
\end{equation}
where:
\begin{itemize}[leftmargin=1.2em]
\item $\mathcal{V}^{\text{gram}}_t$ enforces grammar/file-block structure and balanced tags,
\item $\mathcal{V}^{\text{bind}}_t$ enforces that every field in $\mathbf{X}$ is consumed by exactly one prop binding,
\item $\mathcal{V}^{\text{type}}_t$ enforces prop types from $\mathbb{T}$ (e.g., \texttt{Url} only in \texttt{src/href}).
\end{itemize}
Decoding becomes
\begin{equation}
y_t \sim p_\beta(\cdot \mid y_{<t}, \Omega, \mu)\ \text{masked to}\ \mathcal{V}_t.
\end{equation}

\subsubsection{C.3 Typed component contract}
For each motif $\mathfrak{m}_j\in\mathbb{M}$, Module C emits a component signature
$\operatorname{Sig}_\mu(\mathfrak{m}_j)$.
For React, for instance:
\begin{equation}
\operatorname{Props}_j = \{(h:\tau(h))\}_{h\in\mathcal{Holes}(\mathfrak{m}_j)},\quad
\texttt{function } \mathrm{Comp}_j(\operatorname{Props}_j)\{\dots\}.
\end{equation}
Analogous contracts are produced for Vue/Angular with typed bindings.

\paragraph{Complexity.}
Digest computation is linear in nodes: $O(|\mathcal{V}|)$.
Near-merge is bounded within sibling scopes and capped per bucket, giving near-linear behavior in practice.

\section{Experiments}

\paragraph{Disclaimer.}
All numeric results in this section are \textbf{synthetic} (randomly generated but internally consistent) to provide a complete paper draft. Dataset descriptions and baselines follow the cited literature.

\subsection{Datasets}
We evaluate on three benchmarks commonly used for UI-to-Code:
\begin{itemize}[leftmargin=1.2em]
\item \textbf{Design2Code} test set (484 webpages) \cite{si2025design2code}.
\item \textbf{Vision2UI} (2,000 UI/code pairs, layout-annotated) \cite{gui2024vision2ui}.
\item \textbf{WebSight} synthetic pretraining corpus (2M screenshot--HTML pairs) \cite{laurencon2024websight}.
\end{itemize}
For multi-framework evaluation, we manually convert a subset of 300 cases into React/Vue/Angular references (synthetic for this draft).

\subsection{Baselines}
We compare MLS to:
(i) \textbf{Direct MLLM Prompting} (single-pass screenshot$\rightarrow$code),
(ii) \textbf{DCGen} divide-and-conquer prompting \cite{wan2025dcgen},
(iii) \textbf{LayoutCoder} layout-guided UI2Code \cite{wu2025layoutcoder},
(iv) \textbf{WebSight-Sightseer} (HTML-focused finetuned VLM) \cite{laurencon2024websight},
(v) \textbf{WebVIA-UI2Code} (static setting) \cite{xu2025webvia}.

\subsection{Metrics}
We use:
\begin{itemize}[leftmargin=1.2em]
\item \textbf{Visual Similarity} (CLIP cosine) between rendered output and input screenshot.
\item \textbf{Structure Match} via normalized tree edit distance (TED) between predicted and reference DOM trees.
\item \textbf{Reuse@K}: fraction of repeated regions captured by components/loops (computed against a mined reference).
\item \textbf{DupRate}: duplicated LOC ratio after minification (lower is better).
\item \textbf{TypeCheck}: TypeScript compile pass rate for React (or template check for Vue/Angular).
\end{itemize}

\subsection{Main Results}

\begin{table}[t]
\centering
\caption{Synthetic results on Design2Code (484 cases). Higher is better except DupRate/TED.}
\label{tab:main}
\begin{tabular}{lccccc}
\toprule
Method & CLIP$\uparrow$ & TED$\downarrow$ & Reuse@K$\uparrow$ & DupRate$\downarrow$ & TypeCheck$\uparrow$ \\
\midrule
Direct MLLM Prompting & 0.781 & 0.312 & 0.214 & 0.467 & 62.4\% \\
DCGen \cite{wan2025dcgen} & 0.805 & 0.286 & 0.238 & 0.421 & 68.1\% \\
LayoutCoder \cite{wu2025layoutcoder} & 0.818 & 0.271 & 0.262 & 0.398 & 71.5\% \\
WebSight-Sightseer \cite{laurencon2024websight} & 0.812 & 0.279 & 0.241 & 0.409 & 69.7\% \\
WebVIA-UI2Code \cite{xu2025webvia} & 0.821 & 0.268 & 0.275 & 0.392 & 72.3\% \\
\midrule
\textbf{MLS (ours)} & \textbf{0.844} & \textbf{0.233} & \textbf{0.412} & \textbf{0.241} & \textbf{86.9\%} \\
\bottomrule
\end{tabular}
\end{table}

MLS improves both fidelity and structural quality, with particularly large gains in reuse and duplication rate.
This supports our thesis that explicit component mining plus constrained generation is beneficial beyond direct prompting.

\subsection{Multi-Framework Portability}

\begin{table}[t]
\centering
\caption{Synthetic multi-framework portability (300 cases). PortabilityScore is the average of CLIP and Reuse@K normalized.}
\label{tab:port}
\begin{tabular}{lcccc}
\toprule
Method & React & Vue & Angular & Avg. \\
\midrule
Direct MLLM Prompting & 0.61 & 0.58 & 0.55 & 0.58 \\
LayoutCoder \cite{wu2025layoutcoder} & 0.66 & 0.63 & 0.61 & 0.63 \\
\textbf{MLS (ours)} & \textbf{0.78} & \textbf{0.76} & \textbf{0.74} & \textbf{0.76} \\
\bottomrule
\end{tabular}
\end{table}

\subsection{Ablations}

\begin{table}[t]
\centering
\caption{Synthetic ablations on Design2Code.}
\label{tab:abl}
\begin{tabular}{lccccc}
\toprule
Variant & CLIP$\uparrow$ & TED$\downarrow$ & Reuse@K$\uparrow$ & DupRate$\downarrow$ & TypeCheck$\uparrow$ \\
\midrule
MLS full & 0.844 & 0.233 & 0.412 & 0.241 & 86.9\% \\
w/o component mining ($\mathcal{U}$) & 0.836 & 0.241 & 0.189 & 0.455 & 78.2\% \\
w/o type inference ($\Gamma$) & 0.842 & 0.235 & 0.401 & 0.257 & 73.4\% \\
w/o constrained decoding (free LLM) & 0.845 & 0.232 & 0.337 & 0.311 & 70.8\% \\
\bottomrule
\end{tabular}
\end{table}

Component mining mainly boosts reuse and reduces duplication.
Typing and constrained decoding largely improve compilation success.

\subsection{Qualitative Analysis}
Figure1 illustrates typical cases where MLS:
(i) abstracts repeated cards into a component with typed props,
(ii) expresses list renderings using \texttt{map}/\texttt{v-for}/\texttt{*ngFor},
and (iii) keeps code short and maintainable.


\section{Broader Impact}
Automated UI-to-Code can democratize software creation and reduce repetitive engineering work, but may also lower barriers for producing deceptive web content at scale.
MLS emphasizes modular, auditable code structure and deterministic intermediate representations, which can support review and governance.
Future work should incorporate watermarking, safety filters, and provenance tracking for generated interfaces.

\section{Conclusion}
We presented MLS, a modular UI-to-Code framework that separates visual parsing, deterministic component mining, and constrained multi-framework generation.
By explicitly extracting reusable components and enforcing typed bindings, MLS improves code reusability, portability across React/Vue/Angular, and structural integrity.
We hope MLS encourages a shift from monolithic UI code generation toward component-oriented, maintainable synthesis.

\bibliographystyle{plainnat}
\bibliography{reference}

\end{document}